\documentclass[reprint,aps,showpacs]{revtex4-1}
\usepackage[utf8]{inputenc}
\usepackage[T1]{fontenc}
\usepackage{amsmath}
\usepackage{amsfonts}
\usepackage{amssymb}
\usepackage{graphicx}
\usepackage{bm}
\usepackage{xcolor}
\usepackage[left=2.00cm, right=2.00cm, top=3.00cm, bottom=3.00cm]{geometry}
\begin{document}
	\title{Finite-emittance Wigner crystals in the bubble regime}
	\author{Lars Reichwein}
	\email{lars.reichwein@hhu.de}
	\author{Johannes Thomas}
	\author{Alexander Pukhov}
	\affiliation{Institut f\"{u}r Theoretische Physik I, Heinrich-Heine-Universit\"{a}t D\"{u}sseldorf, D-40225 D\"usseldorf, Germany}
	\date{\today}
	
	\newcommand{\jb}{\textbf{j} }
	\newcommand{\rb}{\textbf{r} }
	\newcommand{\db}{\textbf{d} }
	\newcommand{\nb}{\textbf{n} }
	\newcommand{\pb}{\textbf{p} }
	\newcommand{\vb}{\textbf{v} }
	\newcommand{\Vb}{\textbf{V} }
	\newcommand{\Ab}{\textbf{A} }
	\newcommand{\Bb}{\textbf{B} }
	\newcommand{\Eb}{\textbf{E} }
	\newcommand{\Pb}{\textbf{P} }
	\newcommand{\Fb}{\textbf{F} }
	\newcommand{\Xb}{\textbf{X} }
	\newcommand{\Hb}{\textbf{H} }
	\newcommand{\Rb}{\textbf{R} }
	\newcommand{\jbf}{\textbf{j} }
	\newcommand{\Jbf}{\textbf{J} }
	\newcommand{\kb}{\textbf{k} }
	\newcommand{\eb}{\textbf{e} }
	\newcommand{\ab}{\textbf{a} }
	\newcommand{\xb}{\textbf{x} }
	\newcommand{\yb}{\textbf{y} }
	\newcommand{\qb}{\textbf{q} }
	\newcommand{\Ub}{\textbf{U} }
	\newcommand{\ub}{\textbf{u} }
	\newcommand{\VEC}[3]{\ensuremath{\begin{pmatrix} #1\\ #2\\ #3\end{pmatrix}}}
	\newcommand{\betab}{\bm{\beta}}
	\newcommand{\multiX}{\mathbf{X}}
	\newcommand{\multiV}{\mathbf{V}}
	\newcommand{\multiPi}{\mathbf{\Pi}}
	
	\newcommand{\ext}{\text{ext}}
	\newcommand{\pe}{\text{pe}}
	\newcommand{\phy}{\text{phy}}
	\newcommand{\num}{\text{num}}
	\newcommand{\LW}{\text{LW}}
	\newcommand{\TR}{\text{TR}}
	\newcommand{\tot}{\text{tot}}
	\newcommand{\ret}{\text{ret}}
	\newcommand{\dB}{\text{dB}}
	
	\begin{abstract}
		We study the influence of finite emittance electron bunches in the bubble regime of laser-driven wakefield acceleration onto the microscopic structure of the bunch itself. Using resilient backpropagation (Rprop) to find the equilibrium structure, we observe that for realistic and already observed emittances the previously found crystalline structures remain intact and are only widened marginally. Higher emittances lead to larger electron displacements within the crystal and finally its breaking.
	\end{abstract}

	\maketitle

\section{Introduction}

Laser-driven plasma wakefield acceleration (LWFA) is a promising alternative to conventional accelerators since higher field strengths of more than 100 GV/m can be achieved \cite{Pukhov2002, Malka2012}. Electrons are accelerated using a high-intensity laser (normalized laser amplitude $a_0 > 1$), that is shot into a homogeneous plasma \cite{Tajima1979, Esarey2009}. A solitary electron cavity, the so-called bubble, is created when $a_0  > 4$ and the focal spot size of the laser is twice the Rayleigh length. This bubble is almost spherical and exhibits uniform accelerating fields. During its propagation with almost the speed of light $c$, it can trap electrons from the surrounding plasma (via mechanisms like density down-ramp \cite{Baxevanis2017}, and self-trapping \cite{Leemans2014}), from ionization injection (like in the Trojan Horse regime \cite{Hidding2012a}, or two-color laser injection \cite{Schroeder2018}) or from lateral injection of pre-accelerated bunches \cite{Pronold2018}. Especially interesting are density down-ramp, Trojan Horse and external injection because of the low emittances in the range of $\mu$m mrad and the minimal energy spread of ca. 0.1\% that can be achieved \cite{Hidding2012a, Chen2014, Wang2018, Martinez2017, Tooley2017}.

The macroscopic structure of the electron bunch in the bubble regime has already been studied experimentally \cite{Saevert2015, Schnell2012}, but further inspection of the underlying microscopic structure is also important. The patterning of the beam load is i.a. of interest for the implementation of bright sources of short wavelength radiation.
In conventional Compton backscatter sources a counter-propagating laser pulse is backscattered by a relativistic electron bunch. The resulting photons are of high energy but spatially incoherent, show a wide energy spectrum and are emitted in a broad solid angle \cite{Petrillo2012}. A structuring of the electron bunch would lead to enhances brightness and higher spatial coherence.
In a recent theoretical examination of the microscopic bunch structure electron arrangements similar to Wigner crystals could be observed \cite{Thomas2017}. In the plane transverse to the direction of propagation, these lattices were of hexagonal shape, while longitudinally electronic filaments would form. In this model, a Taylor expansion of the Li\'{e}nard-Wiechert potentials in $v/c$ up to second order was used, thus evading the calculation of the retarded times but therefore neglecting any radiative terms.

In a further publication, the full Li\'{e}nard-Wiechert potentials were considered for 2D slices in the equilibrium slice model (ESM) \cite{Reichwein2018} which allowed for a complete description of the static two-dimensional case and explained the scaling laws for the inter-particle distance width respect to momentum and plasma wavelength analytically. The dependency for the number of particles was determined numerically, since the system of equations describing the equilibrium cannot be solved analytically for a large number of electrons \cite{James1998}. It was shown that smaller inter-particle distances compared to \cite{Thomas2017} are to be expected for the equilibrium structures. 
In \cite{Reichwein2019, Reichwein2020} the 3D structure was examined using a Lorentz transformation which showed electron filaments in the direction of propagation and a formation of various shells around the central filament. These structures were analogous to those observed in typical ring accelerators, however with smaller inter-particle distances in the sub-nanometer regime \cite{Schiffera}.

The two latest approaches still assumed vanishing emittance of the electron bunch, thus the question remains whether the crystalline structures can also be observed for the more realistic case of finite emittance and how comparable the results are to those of \cite{Thomas2017}, which we will be covering in the present work.

In the next section we will give a rough outline of our mathematical model based in the full Li\'{e}nard-Wiechert fields and the corresponding electromagnetic fields. In the scope of this model, we will also discuss how finite emittance influences these fields and therefore the electron-electron interaction. The great-scale effects of emittance will be shown in our section about numerical simulations. 
For different momentum ratios $p_\perp / p_\parallel$ we observe how much the equilibrium structure is changing in contrast to the case of zero emittance and explain this behavior phenomenologically.

\section{Mathematical model}
In the following we will describe the mathematical basis of our model. Our goal is to find a minimum energy structure for a given parameter set, i.e. for given number of electrons $N$ with momenta $p_i$, $i \in \lbrace 1, \dots, N \rbrace$ and given plasma wavelength $\lambda_\pe$. This corresponds to finding a force balance between the external $\Fb_{\ext}$ resulting from the bubble potential (given via the quasi-static 3D bubble model for electron acceleration in homogeneous plasma of \cite{Kostyukov2004}) and the electron-electron interaction for every single electron $\Fb_{C}$ (as shown in \cite{Reichwein2019}):

\begin{align}
	\Fb_i = \Fb_{\ext,i} + \sum_{j=1}^N \Fb_{C, ij} \; .
\end{align}

To calculate the prevailing forces, we need to consider the electric and magnetic fields arising due to the electronic structure. Since we are dealing with relativistic electrons, those fields are given via the  Li\'enard-Wiechert fields

\begin{align}
\Eb(\xb, t) &= e \left[ \frac{\nb - \betab}{\gamma^2 (1 - \betab \cdot \nb)^3 R^2}\right]_\ret \notag \\
& + \frac{e}{c}\left[ \frac{\nb \times ((\nb - \betab) \times  \dot{\betab})}{(1 - \betab \cdot \nb)^3 R}\right]_\ret \; ,\\
\Bb & = [\nb \times \Eb]_\ret \; ,
\end{align}

Here, $e$ is the elementary charge, $\betab = \vb / c$ is the normalized velocity, $\gamma = 1 / \sqrt{1-\beta^2}$ is the Lorentz factor, $R$ is the distance from source to observer and $\nb$ is the corresponding normalized distance vector \cite{Jackson2013}. The index ''ret`` denotes that the variables are given at the retarded time
\begin{align}
	t_\ret = t - \frac{1}{c}|\rb_i - \rb_j (t_\ret)| \; ,
\end{align}
where the observer sits at position $\rb_i$ and receives the signal at time $t$ which is sent at time $t_\ret$ from position $\rb_j$ (the source).

The electric field is split up into two terms, the first being the so-called near field and the second being the far field or radiative field, which is depending on the acceleration $\dot{\betab}$ of the observed particle.

The force acting on each particle due to the electromagnetic fields can be calculated by the Lorentz force
\begin{align}
	\Fb_L = q (\Eb + \vb \times \Bb) \; .
\end{align}

The equilibrium structure corresponds to a positioning of the electrons such that the system's energy is minimized or that the total force onto each particle is vanishing, i.e. $\Fb_i = 0$. This layout of the particles will be found numerically as will be discussed in the following section.

\begin{figure*}[t]
	\includegraphics{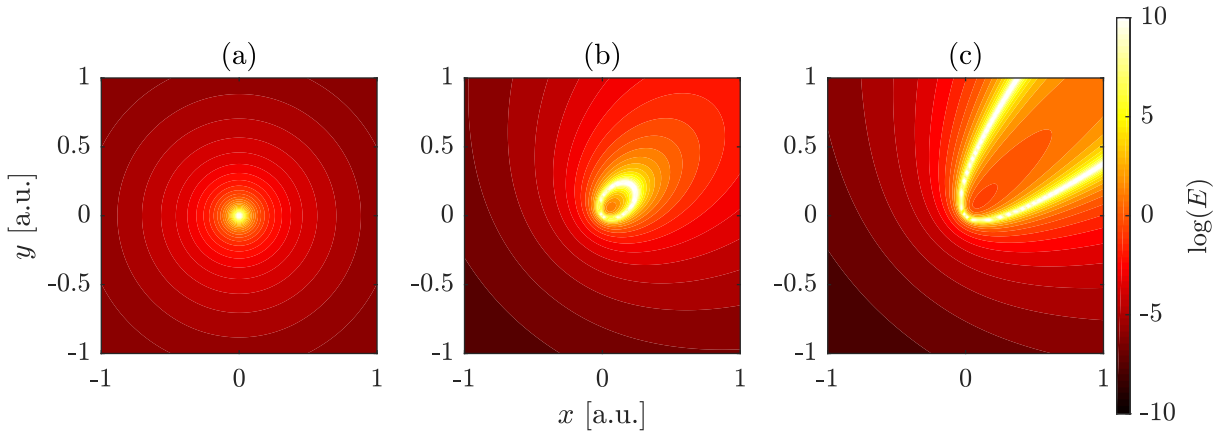}
	\caption{\label{fig:fields}Structure of the electric field in the $x$-$y$-plane created by an electron, sitting in the center of the simulation box. For (a) $p_\perp = 0$, we see a symmetric cone around the propagation direction $\xi$. Increasing the transverse momentum to (b) $p_\perp \approx 0.04 p_\parallel$ and (c)  $p_\perp \approx 0.07 p_\parallel$ leads to a tilt and widening of this cone.}
\end{figure*}

In contrast to the previous publications \cite{Reichwein2018, Reichwein2019, Reichwein2020}, we now consider finite transverse emittance of the electrons. The normalized emittance can be defined as
\begin{align}
	\epsilon = \frac{p_\perp}{p_\parallel} \beta \gamma \delta \; ,
\end{align}
where $p_\perp$ and $p_\parallel$ are the maximum transverse and longitudinal momenta of the particles in the distribution and $\delta$ is the distribution's transverse diameter.
In the following we will confine ourselves to just stating the momentum ratio, i.e. $p_\perp / p_\parallel$, since we are performing simulations with particle numbers far below the actual number of electrons in the bunch, which can have charges in the nC range \cite{Lu2007}. 

We implement the finite emittance in the following way: We consider $N$ electrons, all of which have the same momentum $p_\parallel$ in the direction of propagation.  For the transverse directions, however, we choose a Gaussian distribution of the corresponding momenta, such that for a mean value of $\mu_{p_\perp} = 0$ MeV/c we have a standard deviation of $\sigma_{p_\perp}$, which will be varied in the numerical simulations.

Before going into the specifics of the simulations, we can reason analytically how the electromagnetic fields will be changed by finite emittance.

We consider a simple system of a single electron in the beam load with vanishing transverse emittance. It is accelerated in $\xi$-direction ($\xi = z - V_0 t$ with the bubble velocity $V_0$). The structure of the Li\'{e}nard-Wiechert potentials has the following consequence: information has to travel from this electron to another one in order to see the influence of the resulting electric field. This propagation of information is however limited to the speed of light $c$, thus the field exhibits a cone-like structure (see figure \ref{fig:fields}). If we now consider electrons with transverse emittance, this cone is tilted due to the radial contributions and also widened, leading to a broadening of the whole bunch structure if every single electron has some radial momentum terms.

Going from the case of larger $N$, we see that small transverse momenta only lead to slight changes of the field structure in contrast to zero emittance, while larger transverse terms destroy the field configuration if the transverse momenta are randomly distributed which would lead to the breaking of the crystalline structure.

\section{Numerical simulations}
The simulations for the present paper are performed using the resilient backpropagation (Rprop) algorithm, which will be explained in the following. We start by distributing the $N$ simulated particles randomly in a spherical volume inside the bubble. We then calculate the forces acting on each of the particles. Since we aim for a minimum energy distribution, i.e. an electron configuration where every particle is in a force balance, we try to minimize the occurring forces.

The Rprop algorithm works as follows: Similar to steepest descent we are going against the gradient direction of the function we want to minimize, as the gradient shows in the direction of greatest ascent. As a reminder, the steepest descent algorithm updates the position $\Xb$ via the formula

\begin{align}
	\Xb^{(k + 1)} = \Xb^{(k)} - \alpha^{(k)} \nabla f^{(k)} \; .
\end{align}

Here, $\Xb^{(k)}$ and $\Xb^{(k+1)}$ denote the weights (in our case the electron positions) at iteration $k$ and $k + 1$, $f^{(k)}$ is the function to be minimized evaluated at position $\Xb^{(k)}$. The correct  choice of the step size $\alpha^{(k)}$ is important. If the step size is too small, the algorithm needs many iterations to converge, if it is too large, the algorithm could jump over a minimum. In Rprop, the step size is chosen the following way:

\begin{align}
	\alpha^{(k+1)} = 
	\begin{cases}
		\min(\alpha^{(k)} \eta^{+}, \alpha_{\max}) &\text{if } \nabla f^{(k)} \cdot \nabla f^{(k-1)} > 0 \\
		\max(\alpha^{(k)} \eta^{-}, \alpha_{\min}) &\text{if } \nabla f^{(k)} \cdot \nabla f^{(k-1)} < 0 \\
		\alpha^{(k)}  &\text{else} \\
	\end{cases} \; ,
\end{align}

i.e. the step size is increased if no change of sign is seen in the component observed, otherwise the step size is reduced. The standard values for Rprop are $\eta^{+} = 1.2$, $\eta^{-} = 0.5$, $\alpha_{\max} = 50$ and $\alpha_{\min} = 10^{-6}$  \cite{Riedmillera}, which we have chosen here, too.

We perform several simulations finding the minimum energy structures for various momentum ratios ranging from $p_\perp / p_\parallel = 0$ to $\approx 10^{-2}$. The longitudinal momentum of $p_\parallel = 10$ MeV/c and $\lambda_\pe = 10^{-4}$ m remained fixed throughout.
Assuming a charge of 1 nC for the electron bunch, this would correspond to an emittance range of up to $\epsilon = 2 \times 10^{-5}$ mm mrad.

\begin{figure}
	\includegraphics{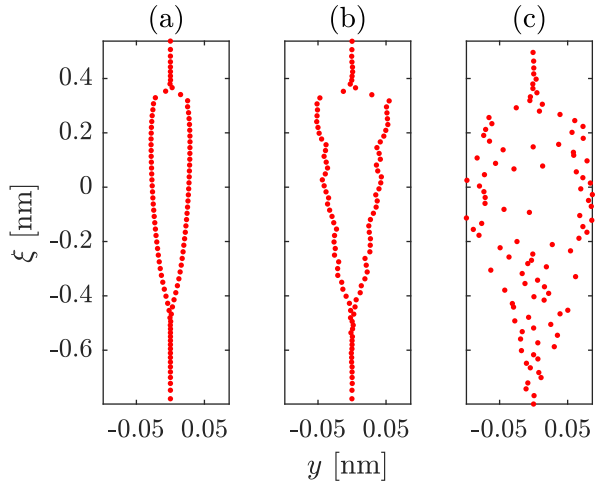}
	\caption{\label{fig:N100}Equilibrium structure of the electron bunch for momentum ratios of (a) $p_\perp \approx 10^{-5} p_\parallel $, (b) $p_\perp \approx 10^{-4} p_\parallel $ and (c) $p_\perp \approx 10^{-2} p_\parallel $. For increasing transverse momentum the structure gets wider and finally breaks apart.}
\end{figure}

\begin{figure}
	\includegraphics{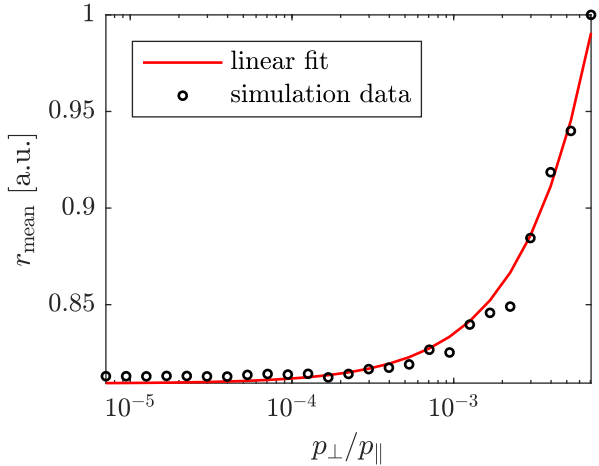}
	\caption{\label{fig:scaling}Increase in the mean radial size of the distribution for larger momentum ratios $p_\perp / p_\parallel$ and corresponding linear fit. The mean radius is normalized to the maximum radius observed for this simulation series.}
\end{figure}

Exemplary, the influence of finite emittance onto the structure is shown for $N = 100$ electrons in figure \ref{fig:N100}. For an ultra-low emittance of  $p_\perp \approx 10^{-5} p_\parallel $ (frame (a)), we see the previously observed central filament that is split into two halves. The splitting is due to the choice of longitudinal momentum $p_\parallel$ and plasma wavelength $\lambda_\pe$ (which is discussed further in \cite{Reichwein2019}). If we now increase the transverse momentum to $p_\perp \approx 10^{-4} p_\parallel $, we see that the structure begins to get corrugated and widened as we expected from our look into the structure of the electromagnetic fields. Further increase of transverse momentum to $p_\perp \approx 10^{-2} p_\parallel $ finally leads to the breaking of the crystalline structure: we still see an elongated shape (which is due to the acceleration in $\xi$-direction), but no real order as in the previous frames.
The broadening of the structure is linear in the transverse momentum, as can be seen in figure \ref{fig:scaling}.

Breaking the crystalline structure would diminish the aforementioned applications of the electron bunch for high-brightness gamma sources, however, the lower emittance values, for which the structure holds, are already experimentally achievable \cite{Hidding2012a, Chen2014}. Furthermore, the effects of transverse emittance are compensated to some extent by a larger number of particles as the random distribution of the radial momenta averages out and the particles can arrange themselves in a way that is close to the case of ultralow emittance (not shown here). For a large enough number of electrons different shells could already be seen in \cite{Reichwein2019}. In this case, the outer shells would press the inner ones together, thus suppressing some effects of the finite emittance that would otherwise lead to breaking.

The widening of the structure due to finite emittance is an effect similar to changing one of the other parameters like longitudinal momentum $p_\parallel$, plasma wavelength $\lambda_\pe$ or particle number $N$. Changing one of those parameters leads to either an increase or a decrease in inter-particle distance, since the force balance between repulsive Coulomb interaction and the focusing bubble potential is shifted in one or the other direction. Similarly, increasing the transverse momentum leads to more repelling between the single electrons. Therefore, not all particles can fit into the central filament and  ''escape`` to places farther from the $\xi$-axis.  Since the transverse momenta are randomly distributed, this pushing outwards is not happening uniformly as it was in the case of zero emittance but instead in a wavy pattern (as seen in figure \ref{fig:N100}).

\section{Conclusion}
In this paper we have simulated the effects of finite emittance onto the electron bunch of the bubble regime of LWFA. We have modeled the interaction of the electrons with the bubble fields and with each other using the full Li\'{e}nard-Wiechert potentials.  In a approach similar to the previous descriptions of the bunch structure \cite{Thomas2017, Reichwein2018, Reichwein2019, Reichwein2020}, we found its equilibrium by distributing the electrons randomly inside a unit sphere and iteratively minimizing the system's energy, here using Rprop.
For small transverse momenta, the structure of the bunch remains largely unchanged in comparison to the zero-emittance case only showing some smaller corrugations and a slight widening of the total structure. If the emittance is high enough, the crystal can break apart. The emittance values for which the lattice still holds are however experimentally achievable \cite{Hidding2012a, Chen2014}.

\begin{acknowledgments}

This work has been supported in parts by DFG (projects PU 213/6-1 and PU 213/9-1) and by BMBF (project 05K19PFA).
\end{acknowledgments}

\bibliography{ref_plasma}
\end{document}